\documentclass[nofootinbib,prd,showpacs]{revtex4}%
\usepackage{amsmath}
\usepackage{amsfonts}
\usepackage{amssymb}
\usepackage{graphicx}%
\begin{document}
\title{Self sustained phantom wormholes in semi-classical gravity}
\author{Remo Garattini}
\email{Remo.Garattini@unibg.it}
\affiliation{Universit\`{a} degli Studi di Bergamo, Facolt\`{a} di Ingegneria, Viale
Marconi 5, 24044 Dalmine (Bergamo) ITALY}
\affiliation{INFN - sezione di Milano, Via Celoria 16, Milan, Italy.}
\author{Francisco S. N. Lobo}
\email{flobo@cosmo.fis.fc.ul.pt} \affiliation{Institute of
Gravitation \& Cosmology, University of Portsmouth, Portsmouth PO1
2EG, UK} \affiliation{Centro de Astronomia e Astrof\'{\i}sica da
Universidade de Lisboa, Campo Grande, Ed. C8 1749-016 Lisboa,
Portugal}

\begin{abstract}
A possible candidate for the late time accelerated expanding Universe is
phantom energy, which possesses rather bizarre properties, such as the
prediction of a Big Rip singularity and the violation of the null energy
condition. The latter is a fundamental ingredient of traversable wormholes,
and it has been shown that phantom energy may indeed sustain these exotic
geometries. Inspired by the evolving dark energy parameter crossing the
phantom divide, we consider in this work a varying equation of state parameter
dependent on the radial coordinate, i.e., $\omega(r)=p(r)/\rho(r)$. We shall
impose that phantom energy is concentrated in the neighborhood of the throat,
to ensure the flaring out condition, and several models are analyzed. We shall
also consider the possibility that these phantom wormholes be sustained by
their own quantum fluctuations. The energy density of the graviton one loop
contribution to a classical energy in a phantom wormhole background and the
finite one loop energy density are considered as a self-consistent source for
these wormhole geometries. The latter semi-classical approach prohibits
solutions with a constant equation of state parameter, which further motivates
the imposition of a radial dependent parameter, $\omega(r)$, and only permits
solutions with a steep positive slope proportional to the radial derivative of
the equation of state parameter, evaluated at the throat. The size of the
wormhole throat as a function of the relevant parameters is also explored.

\end{abstract}

\maketitle

\section{Introduction}

Unravelling the source for the late time accelerated expansion of
the Universe has become a central theme in modern cosmology
\cite{observations}. Several candidates, responsible for this
expansion, have been proposed in the literature, ranging from
modified gravity theories to the introduction of cosmological
models comprising of a negative pressure cosmic fluid, denoted as
dark energy. The latter is parameterized by the equation of state
$\omega=p/\rho$, where $p$ is the spatially homogeneous pressure
and $\rho$ the dark energy density. A value of $\omega<-1/3$ is
required to ensure the cosmic accelerated expansion, and
$\omega=-1$ corresponds to the presence of a cosmological
constant. It is interesting that recent constraints from
observational data have confirmed that the dark energy parameter
lies slightly below the characteristic cosmological constant
value, $\omega=-1$ \cite{3yWMAP}. Ameliorated fits to the data
also indicate that an evolving equation of state parameter which
crosses $\omega=-1$ is mildly favored \cite{phantom-divide}. The
models satisfying $\omega<-1$ are denoted as \textit{phantom
energy} models, and possess rather interesting characteristics, in
particular, the prediction of the \textit{Big Rip} singularity and
the violation of the null energy condition (NEC). The latter is a
fundamental ingredient of traversable wormholes, and it was
recently shown that phantom energy may indeed support these exotic
geometries \cite{phantomWH,phantomWH2}. Inspired by the evolving
dark energy parameter crossing the phantom divide, we consider in
this work a varying equation of state parameter dependent on the
radial coordinate, i.e., $\omega(r)=p(r)/\rho(r)$, and impose that
phantom energy is concentrated in the neighborhood of the throat,
in order to ensure the NEC violation, $\omega(r_{0})<-1$, where
$r_{0}$ is the wormhole throat radius. The latter restriction may
be relaxed far from the throat. Despite the fact that the dark
(phantom) energy equation of state represents a spatially
homogeneous cosmic fluid, due to gravitational instabilities
inhomogeneities may arise.
Note that now one regards that the pressure in the equation of
state $p=\omega \rho$ is a negative radial pressure, and the
tangential pressure $p_t$ is determined through the Einstein field
equations. This is fundamentally the analysis carried out in Ref.
\cite{Sush-Kim}, where the authors constructed a time-dependent
solution describing a spherically symmetric wormhole in a
cosmological setting with a ghost scalar field. It was shown that
the radial pressure is negative everywhere and far from the
wormhole throat equals the transverse pressure, showing that the
ghost scalar field behaves essentially as dark energy.
Another interesting aspect of this topic is the effect of phantom
energy accretion in wormholes immersed in this cosmic fluid. In
fact, Gonz\'{a}lez-D\'{i}az analyzed the evolution of wormhole and
ringhole spacetimes embedded in a background accelerating
Universe~\cite{gonzalez2} driven by dark energy, and further
considered the accretion of dark and phantom energy onto
Morris-Thorne wormholes~\cite{diaz-phantom3}. It was shown that
this accretion gradually increases the wormhole throat which
eventually overtakes the accelerated expansion of the universe and
becomes infinite at a time in the future before the big rip.

Thus, the phantom wormhole models considered in this work may
possibly arise from density fluctuations in the cosmological
background \cite{VDW}. In this context, we shall also be
interested in the analysis that these phantom wormholes be
sustained by their own quantum fluctuations. As noted above,
wormholes violate the NEC, and consequently violate all of the
other classical energy conditions. Thus, it seems that these
exotic spacetimes arise naturally in the quantum regime, as a
large number of quantum systems have been shown to violate the
energy conditions, such as the Casimir effect. Indeed, various
wormhole solutions in semi-classical gravity have been considered
in the literature. For instance, semi-classical wormholes were
found in the framework of the Frolov-Zelnikov approximation for
$\langle T_{\mu\nu}\rangle$ \cite{Sushkov}. Analytical
approximations of the stress-energy tensor of quantized fields in
static and spherically symmetric wormhole spacetimes were also
explored in Refs. \cite{Popov}. However, the first self-consistent
wormhole solution coupled to a quantum scalar field was obtained
in Ref. \cite{Hochberg}. The ground state of a massive scalar
field with a non-conformal coupling on a short-throat flat-space
wormhole background was computed in Ref. \cite{Khusn}, by using a
zeta renormalization approach. The latter wormhole model, which
was further used in the context of the Casimir effect \cite{Khab},
was constructed by excising spherical regions from two identical
copies of Minkowski spacetime, and finally surgically grafting the
boundaries (A more realistic geometry was considered in Ref.
\cite{Khusn2}). In this context, this motivates the semi-classical
analysis in this work, where the Einstein field equation takes the
form
\begin{equation}
G_{\mu\nu}=\kappa\,\langle T_{\mu\nu}\rangle^{\mathrm{ren}}\,,
\end{equation}
with $\kappa=8\pi G$, and $\langle T_{\mu\nu}\rangle^{\mathrm{ren}}$ is the
renormalized expectation value of the stress-energy tensor operator of the
quantized field. The formalism outlined in this work follows Ref.
\cite{Garattini} very closely. Now, the metric may be separated into a
background component, $\bar{g}_{\mu\nu}$ and a perturbation $h_{\mu\nu}$,
i.e., $g_{\mu\nu}=\bar{g}_{\mu\nu}+h_{\mu\nu}$. The Einstein tensor may also
be separated into a part describing the curvature due to the background
geometry and that due to the perturbation, i.e.,
\begin{equation}
G_{\mu\nu}(g_{\alpha\beta})=G_{\mu\nu}(\bar{g}_{\alpha\beta})+\Delta G_{\mu
\nu}(\bar{g}_{\alpha\beta},h_{\alpha\beta})\,,
\end{equation}
where $\Delta G_{\mu\nu}(\bar{g}_{\alpha\beta},h_{\alpha\beta})$ may be
considered a perturbation series in terms of $h_{\mu\nu}$. Using the
semi-classical Einstein field equation, in the absence of matter fields, one
may define an effective stress-energy tensor for the quantum fluctuations as
\begin{equation}
\langle T_{\mu\nu}\rangle^{\mathrm{ren}}=-\frac{1}{\kappa}\,\langle\Delta
G_{\mu\nu}(\bar{g}_{\alpha\beta})\rangle^{\mathrm{ren}}\,,
\end{equation}
so that the equation governing quantum fluctuations behaves as a backreaction
equation. The semi-classical procedure followed in this work relies heavily on
the formalism outlined in Ref. \cite{Garattini}, where the graviton one loop
contribution to a classical energy in a traversable wormhole background was
computed, through a variational approach with Gaussian trial wave functionals
\cite{Garattini,Garattini2}. A zeta function regularization is used to deal
with the divergences, and a renormalization procedure is introduced, where the
finite one loop is considered as a self-consistent source for traversable
wormholes. Rather than reproduce the formalism, we shall refer the reader to
Ref. \cite{Garattini} for details, when necessary. In this paper, rather than
integrate over the whole space as in Ref.\cite{Garattini}, we shall work with
the energy densities, which provides a more general working hypothesis.

This paper is outlined in the following manner: In Section \ref{sec:field}, we
outline the field equations governing phantom wormholes with an $r$-dependent
parameter, $\omega(r)$, and explore specific solutions. In Section
\ref{sec:self}, we analyze self-sustained traversable phantom wormholes in
semi-classical gravity, considering that the equation governing quantum
fluctuations behaves as a backreaction equation. Finally, in Section
\ref{sec:conlusion}, we conclude.

\section{Field equations and solutions}

\label{sec:field}

The spacetime metric representing a spherically symmetric and static wormhole
is given by
\begin{equation}
ds^{2}=-e ^{2\Phi(r)}\,dt^{2}+\frac{dr^{2}}{1- b(r)/r}+r^{2} \,(d\theta
^{2}+\sin^{2}{\theta} \, d\phi^{2}) \,, \label{metricwormhole}%
\end{equation}
where $\Phi(r)$ and $b(r)$ are arbitrary functions of the radial coordinate,
$r$, denoted as the redshift function, and the form function, respectively
\cite{Morris}. The radial coordinate has a range that increases from a minimum
value at $r_{0}$, corresponding to the wormhole throat, to infinity. A
fundamental property of a wormhole is that a flaring out condition of the
throat, given by $(b-b^{\prime}r)/b^{2}>0$, is imposed \cite{Morris,Visser},
and at the throat $b(r_{0})=r=r_{0}$, the condition $b^{\prime}(r_{0})<1$ is
imposed to have wormhole solutions. Another condition that needs to be
satisfied is $1-b(r)/r>0$. For the wormhole to be traversable, one must demand
that there are no horizons present, which are identified as the surfaces with
$e^{2\Phi}\rightarrow0$, so that $\Phi(r)$ must be finite everywhere.

Using the Einstein field equation, $G_{\mu\nu}=\kappa\,T_{\mu\nu}$, with
$\kappa=8\pi G$ and $c=1$, we obtain the following relationships
\begin{align}
b^{\prime}  &  =\kappa\, r^{2} \rho\,,\label{rhoWH}\\
\Phi^{\prime}  &  =\frac{b+\kappa\,r^{3} p_{r}}{2r^{2}(1-b/r)} \,,\label{prWH}%
\\
p_{r}^{\prime}  &  =\frac{2}{r}\,(p_{t}-p_{r})-(\rho+p_{r})\,\Phi^{\prime}\,,
\label{ptWH}%
\end{align}
where the prime denotes a derivative with respect to the radial coordinate,
$r$. $\rho(r)$ is the energy density, $p_{r}(r)$ is the radial pressure, and
$p_{t}(r)$ is the lateral pressure measured in the orthogonal direction to the
radial direction.

Now consider the equation of state representing phantom energy,
but with a varying parameter $\omega(r)$, given by
$p_{r}(r)=\omega(r) \rho(r)$ with $\omega<-1$ in the phantom
regime \cite{phantomWH,phantomWH2}. The motivation for this
analysis arises from a time-dependent crossing of the phantom
divide, however, here we shall consider an $r$-dependent equation
of state, with the imposition of phantom energy distribution in
the throat neighborhood, with $\omega (r_{0})<-1$, to ensure the
NEC violation, and consequently the flaring-out condition. This
condition may be relaxed far from the throat. Taking into account
Eqs. (\ref{rhoWH})-(\ref{prWH}), we have the following condition
\begin{equation}
\Phi^{\prime}(r)=\frac{b(r)+\omega(r) rb^{\prime}(r)}{2r^{2}\,\left[  1-b(r)/r
\right]  } \,. \label{EOScondition}%
\end{equation}
We now have four equations, namely the field equations, i.e., Eqs.
(\ref{rhoWH})-(\ref{ptWH}), and Eq. (\ref{EOScondition}), with six unknown
functions of $r$, i.e., $\rho(r)$, $p_{r}(r)$, $p_{t}(r)$, $b(r)$, $\Phi(r)$
and $\omega(r)$. One may adopt different strategies to construct solutions
with the properties and characteristics of wormholes. One may consider a
plausible matter-field distribution, by considering convenient stress-energy
tensor components, and through the field equations and Eq. (\ref{EOScondition}%
) determine the metric fields and the function $\omega(r)$.

Alternatively, by appropriately choosing the form function and $\omega(r)$,
one may integrate Eq. (\ref{EOScondition}), to determine the redshift
function, $\Phi(r)$. As an example, consider the following functions
\begin{align}
b(r)  &  =r_{0}+\gamma^{2}r_{0}\left(  1-\frac{r_{0}}{r}\right)\,, \label{f1}\\
\omega(r)  &  =-\alpha^{2}+\beta^{2}\left(  1-\frac{r_{0}}{r}\right)  \,,
\label{w1}%
\end{align}
with $0<\gamma^{2}<1$, so that one obtains a wormhole solution with
$b^{\prime}(r)=\gamma^{2}r_{0}^{2}/r^{2}$, and at the throat we have
$b(r_{0})=r_{0}$ and $b^{\prime}(r_{0})<1$. Relatively to the function
$\omega(r)$, we impose $\omega(r_{0})=-\alpha^{2}<-1$, and $-1<\omega
(r)=-\alpha^{2}+\beta^{2}<-1/3$ as $r\rightarrow\infty$. Now, substituting
Eqs. (\ref{f1})-(\ref{w1}) into Eq. (\ref{EOScondition}), we obtain the
solution
\begin{align}
\Phi(r)  &  =-\frac{1-\gamma^{2}\alpha^{2}}{2(\gamma^{2}-1)}\,\ln(r-r_{0})
-\frac{1}{2}\left(  1+\alpha^{2}+\frac{\beta^{2}}{\gamma^{2}}\right)
\ln(r)\nonumber\\
&  \hspace{-0.5cm}+\frac{\beta^{2}-\alpha^{2}-\frac{\beta^{2}} {\gamma^{2}%
}+\gamma^{2}}{2(\gamma^{2}-1)}\,\ln(r-\gamma^{2}r_{0}) +\frac{\beta^{2}r_{0}%
}{2r}+C \,.
\end{align}
This corresponds to a non-traversable wormhole solution, due to the presence
of an event horizon, as may be readily verified from the first term in the
right hand side. Therefore, we fine-tune the parameter $\alpha^{2}%
=1/\gamma^{2}$, so that the redshift function finally takes the form
\begin{align}
\Phi(r)=\frac{1+\beta^{2}+\gamma^{2}}{2\gamma^{2}}\,\ln\left(  1-\frac
{\gamma^{2}r_{0}}{r}\right)  +\frac{\beta^{2}r_{0}}{2r} \,,
\end{align}
which is now finite throughout the range of $r$, and where we have considered
$C=0$, without a significant loss of generality. Note that this solution
corresponds to an asymptotically flat traversable wormhole, as $b(r)/r
\rightarrow0$ and $\Phi(r)\rightarrow0$ for $r \rightarrow\infty$.

One may also consider restricted choices for $\Phi(r)$ and $\omega(r)$, and
through Eq. (\ref{EOScondition}) determine the form function through the
following relationship
\begin{equation}
b(r)=e^{-\Gamma(r_{0},r)}\,\left[  r_{0}+\int_{r_{0}}^{r}\frac{2\Phi^{\prime
}(\bar{r})\bar{r}} {\omega(\bar{r})\bar{r}}\;e^{\Gamma(r_{0},\bar{r})}%
\,d\bar{r}\right]  \,, \label{diffeq}%
\end{equation}
with
\begin{equation}
\Gamma(r_{0},r)=\int_{r_{0}}^{r}\frac{2\Phi^{\prime}(\bar{r})\bar{r}+1}
{\omega(\bar{r})\bar{r}}\,d\bar{r}\,.
\end{equation}
Now, interestingly enough results may be inferred by imposing a constant
redshift function, $\Phi^{\prime}=0$, i.e., $b(r)+\omega(r)rb^{\prime}(r)=0$,
which may be integrated to yield the following form function
\begin{equation}
b(r)=r_{0}\,\exp\left[  -\int_{r_{0}}^{r}\,\frac{d\bar{r}}{\omega(\bar{r}%
)\bar{r}}\right]  \,. \label{form}%
\end{equation}
The latter may also be deduced from Eq. (\ref{diffeq}), by setting
$\Phi^{\prime}(r)=0$. Note that for a constant equation of state
parameter, $\omega$, the form function is given by
$b(r)=r_0(r/r_0)^{-1/\omega}$, which is a solution analyzed in
detail in Ref. \cite{phantomWH2}.

For instance, consider the following function
\begin{equation}
\label{w2}\omega(r)=\frac{\alpha^{2}\beta^{2}(r/r_{0})}{\alpha^{2}\left(
1-r/r_{0}\right)  -\beta^{2}}\,,
\end{equation}
which takes the values $\omega(r_{0})=-\alpha^{2}<-1$ at the throat, and
$\omega(r)=-\beta^{2}$ as $r\rightarrow\infty$. Now, substituting this
specific function in Eq. (\ref{form}), we deduce the following form function
\begin{equation}
b(r)=r_{0}\,\left(  \frac{r}{r_{0}}\right)  ^{1/\beta^{2}}\, \exp\left[
-\left(  1-\frac{r_{0}}{r}\right)  \left(  \frac{1}{\beta^{2}} -\frac
{1}{\alpha^{2}}\right)  \right]  \,. \label{f2}%
\end{equation}
Note that if we wish to impose the crossing of the phantom divide the
condition $\beta^{2}<1$ needs to be satisfied. However, from the expression of
the form function, we verify that in order to have an asymptotically flat
spacetime, i..e, $b/r\rightarrow0$, as $r\rightarrow\infty$, the condition
$\beta^{2}>1$ is imposed. Therefore for this particular choice of $\omega(r)$,
the solution remains in the phantom regime for all values of $r$, so that the
wormhole is constituted totally of phantom energy. The derivative of the form
function evaluated at the throat, is given by $b^{\prime}(r_{0})=1/\alpha
^{2}<1$, consequently satisfying the flaring-out condition.

Now, consider the function given by Eq. (\ref{w1}), then integrating Eq.
(\ref{form}), we deduce the following form function
\begin{equation}
b(r)=r_{0} \left[  \frac{\beta^{2}}{\alpha^{2}}+\frac{r_{0}}{r} \left(
1-\frac{\beta^{2}}{\alpha^{2}}\right)  \right]  ^{1/(\alpha^{2}-\beta^{2})}
\,. \label{f3}%
\end{equation}
Note that for this case to be an asymptotically flat spacetime, then we need
to impose $\alpha^{2}-\beta^{2}>1$. This latter restriction imposes that
$\omega$ remains in the phantom regime, $\omega(r)<-1$ for all $r$. The
derivative of the form function, evaluated at the throat, is given by
$b^{\prime}(r_{0})=1/\alpha^{2}<1$, which also satisfies the flaring out
condition. We note that it is possible, in principle, to construct a wide
variety of traversable wormholes supported by phantom energy, with an
$r$-dependent equation of state parameter.

\section{Self-sustained phantom wormholes}

\label{sec:self}

In this Section we shall consider the formalism outlined in detail in Ref.
\cite{Garattini}, where the graviton one loop contribution to a classical
energy in a wormhole background is used. The latter contribution is evaluated
through a variational approach with Gaussian trial wave functionals, and the
divergences are treated with a zeta function regularization. Using a
renormalization procedure, the finite one loop energy was considered a
self-consistent source for a traversable wormhole. We refer the reader to Ref.
\cite{Garattini} for details.

The classical energy is given by
\[
H_{\Sigma}^{(0)}=\int_{\Sigma}\,d^{3}x\,\mathcal{H}^{(0)}=-\frac{1}{16\pi
G}\int_{\Sigma}\,d^{3}x\,\sqrt{g}\,R\,,
\]
where the background field super-hamiltonian, $\mathcal{H}^{(0)}$, is
integrated on a constant time hypersurface. $R$ is the curvature scalar, and
using metric (\ref{metricwormhole}), is given by
\[
R=-2\left(  1-\frac{b}{r}\right)  \left[  \Phi^{\prime\prime}+(\Phi^{\prime
})^{2}-\frac{b^{\prime}}{r(r-b)}-\frac{b^{\prime}r+3b-4r}{2r(r-b)}%
\,\Phi^{\prime}\right]  \,.
\]
We shall henceforth consider a constant redshift function, $\Phi^{\prime
}(r)=0$, which provides interestingly enough results, so that the curvature
scalar reduces to $R=2b^{\prime}/r^{2}$. Thus, the classical energy reduces
to
\begin{equation}
H_{\Sigma}^{(0)}=\frac{1}{2G}\int_{r_{0}}^{\infty}\,\frac{dr\,r^{2}}%
{\sqrt{1-b(r)/r}}\,\frac{b(r)}{\omega(r)r^{3}}\,, \label{classical}%
\end{equation}
where the condition $b^{\prime}(r)=-b(r)/[\omega(r)r]$ has been used to
eliminate the $b^{\prime}(r)$ dependence.

We shall also take into account the total regularized one loop energy given
by
\begin{equation}
E^{TT}= 2\int_{r_{0}}^{\infty}\,dr\frac{r^{2}}{\sqrt{1-b(r)/r}}\,\left[
\rho_{1}(\varepsilon)+\rho_{2}(\varepsilon)\right]  \,,
\end{equation}
where once again, we refer the reader to Ref. \cite{Garattini} for details.
The energy densities, $\rho_{i}(\varepsilon)$ (with $i=1,2$), are defined as
\begin{equation}
\rho_{i}(\varepsilon)=\frac{1}{4\pi}\mu^{2\varepsilon}\int_{\sqrt{U_{i}(r)}%
}^{\infty}\,d\tilde{E}_{i}\,\frac{\tilde{E}_{i}^{2}}{\left[  \tilde{E}_{i}%
^{2}-U_{i}(r)\right]  ^{\varepsilon-1/2}}=-\frac{U_{i}^{2}(r)}{64\pi^{2}%
}\left[  \frac{1}{\varepsilon}+\ln\left(  \frac{\mu^{2}}{U_{i}}\right)
+2\ln2-\frac{1}{2}\right]  . \label{energy}%
\end{equation}
The zeta function regularization method has been used to determine the energy
densities, $\rho_{i}$. It is interesting to note that this method is identical
to the subtraction procedure of the Casimir energy computation, where the zero
point energy in different backgrounds with the same asymptotic properties is
involved. In this context, the additional mass parameter $\mu$ has been
introduced to restore the correct dimension for the regularized quantities.
Note that this arbitrary mass scale appears in any regularization scheme.

Note that in analyzing the Einstein field equations, one should
consider the whole system of equations, which in the static
spherically symmetric case also includes the $rr$-component. The
joint analysis of the equations is necessary to guarantee the
compatibility of the system. However, in the semi-classical
framework considered in this work there is no dynamical equation
for the pressure. Nevertheless, one may argue that the
semi-classical part of the pressure is known through the equation
of state that determines the relation between the energy density
and the pressure.

Since a self sustained wormhole must satisfy%
\begin{equation}
H_{\Sigma}^{(0)}=-\,E^{TT},
\end{equation}
which is an integral relation, this has to be true also for the integrand,
namely the energy density. Therefore, we set%
\begin{equation}
\frac{1}{2G}\frac{b(r)}{\omega(r)r^{3}}=-2\,\left[  \rho_{1}(\varepsilon
)+\rho_{2}(\varepsilon)\right]  \,. \label{Ett}%
\end{equation}
At first sight one can think that the \textit{phantom} region is
straightforward to analyze. However, within the approach of Ref.
\cite{Garattini}, this is not trivial as the main difference is in the
integration. In this paper, we shall consider the approach of working with the
energy density, while in Ref. \cite{Garattini}, the integration over the whole
space was considered. In Appendix \ref{app}, we show that the classical term
integrated over all space forbids the appearance of \textit{phantom
energy}~\cite{Garattini1}, at least for constant $\omega$.

For this purpose, the Lichnerowicz equations provide the potentials, which are
given by
\begin{align}
U_{1}(r)  &  =\frac{6}{r^{2}}\,\left[  1-\frac{b(r)}{r}\right]  -\frac
{3}{2r^{2}}\,\left[  b^{\prime}(r)-\frac{b(r)}{r}\right] \nonumber\\
&  =\frac{6}{r^{2}}\,\left[  1-\frac{b(r)}{r}\right]  +\frac{3b(r)}{2r^{3}%
}\,\left[  \frac{1}{\omega(r)}+1\right]  \,, \label{p1}%
\end{align}%
\begin{align}
U_{2}(r)  &  =\frac{6}{r^{2}}\,\left[  1-\frac{b(r)}{r}\right]  -\frac
{3}{2r^{2}}\,\left[  \frac{-b^{\prime}(r)}{2}+\frac{b(r)}{r}\right]
\nonumber\\
&  =\frac{6}{r^{2}}\,\left[  1-\frac{b(r)}{r}\right]  +\frac{3b(r)}{2r^{3}%
}\,\left[  \frac{1}{3\omega(r)}-1\right]  \,, \label{p2}%
\end{align}
where the relationship $b^{\prime}(r)=-b(r)/[\omega(r)r]$ has been used once
again in the last terms, to eliminate the $b^{\prime}(r)$ dependence. We refer
the reader to Ref. \cite{Garattini} for the deduction of these expressions.

Thus, taking into account Eq. (\ref{energy}), then Eq. (\ref{Ett}) yields the
following relationship%
\begin{equation}
\hspace{-1.5cm}\frac{1}{2G}\,\frac{b(r)}{\omega(r)r^{3}}=\frac{1}{32\pi
^{2}\varepsilon}\,\left[  U_{1}^{2}(r)+U_{2}^{2}(r)\right]  +\frac{1}%
{32\pi^{2}}\left[  U_{1}^{2}\ln\left(  \left\vert \frac{4\mu^{2}}{U_{1}%
\sqrt{e}}\right\vert \right)  +U_{2}^{2}\ln\left(  \left\vert \frac{4\mu^{2}%
}{U_{2}\sqrt{e}}\right\vert \right)  \right]  . \label{selfsust}%
\end{equation}
It is essential to renormalize the divergent energy by absorbing the
singularity in the classical quantity, by redefining the bare classical
constant $G$ as
\begin{equation}
\frac{1}{G}\rightarrow\frac{1}{G_{0}}+\frac{2}{\varepsilon}\,\frac{\left[
U_{1}^{2}(r)+U_{2}^{2}(r)\right]  }{32\pi^{2}}\frac{\omega(r)r^{3}}{b(r)}\,.
\end{equation}
Using this, Eq. (\ref{selfsust}) takes the form
\begin{equation}
\frac{1}{2G_{0}}\,\frac{b(r)}{\omega(r)r^{3}}=\frac{1}{32\pi^{2}}\left[
U_{1}^{2}\ln\left(  \left\vert \frac{4\mu^{2}}{U_{1}\sqrt{e}}\right\vert
\right)  +U_{2}^{2}\ln\left(  \left\vert \frac{4\mu^{2}}{U_{2}\sqrt{e}%
}\right\vert \right)  \right]  . \label{selfsust2}%
\end{equation}
Note that this quantity depends on an arbitrary mass scale. Thus, using the
renormalization group equation to eliminate this dependence, we impose that
\begin{equation}
\mu\frac{d}{d\mu}\left[  \frac{1}{2G_{0}}\,\frac{b(r)}{\omega(r)r^{3}}\right]
=\frac{\mu}{32\pi^{2}}\frac{d}{d\mu}\left[  U_{1}^{2}\ln\left(  \left\vert
\frac{4\mu^{2}}{U_{1}\sqrt{e}}\right\vert \right)  +U_{2}^{2}\ln\left(
\left\vert \frac{4\mu^{2}}{U_{2}\sqrt{e}}\right\vert \right)  \right]  \,,
\end{equation}
which reduces to
\begin{equation}
\frac{b(r)}{\omega(r)r^{3}}\,\mu\,\frac{\partial G_{0}^{-1}(\mu)}{\partial\mu
}=\frac{1}{8\pi^{2}}\,\left[  U_{1}^{2}(r)+U_{2}^{2}(r)\right]  \,.
\label{selfsust3}%
\end{equation}
The renormalized constant $G_{0}$ is treated as a running constant, in the
sense that it varies provided that the scale $\mu$ is varying, so that one may
consider the following definition
\begin{equation}
\frac{1}{G_{0}(\mu)}=\frac{1}{G_{0}(\mu_{0})}+\frac{\left[  U_{1}^{2}%
(r)+U_{2}^{2}(r)\right]  }{8\pi^{2}}\,\,\frac{\omega(r)r^{3}}{b(r)}%
\,\ln\left(  \frac{\mu}{\mu_{0}}\right)  \,. \label{selfsust4a}%
\end{equation}
Thus, Eq. (\ref{selfsust}) finally provides us with
\begin{equation}
\frac{1}{G_{0}(\mu_{0})}=\frac{\omega(r)r^{3}}{16\pi^{2}b(r)}\,\left[
U_{1}^{2}(r)\ln\left(  \left\vert \frac{4\mu_{0}^{2}}{U_{1}(r)\sqrt{e}%
}\right\vert \right)  +U_{2}^{2}(r)\ln\left(  \left\vert \frac{4\mu_{0}^{2}%
}{U_{2}(r)\sqrt{e}}\right\vert \right)  \right]  . \label{selfsust4}%
\end{equation}
Now, the procedure that we shall follow is to find the extremum of the right
hand side of Eq. (\ref{selfsust4}) with respect to $r$, and finally evaluate
at the throat $r_{0}$, in order to have only one solution (see discussion in
Ref. \cite{Garattini}).

To this effect, we shall use the derivative of the potentials,
Eqs. (\ref{p1})-(\ref{p2}), which we write down for
self-completeness and self-consistency, and are given by
\begin{align}
U_{1}^{\prime}(r)  &  =-\frac{12}{r^{3}}\left(  1-\frac{b}{r}\right)
+\frac{6(b-b^{\prime}r)}{r^{4}} +\frac{3}{2}\left[  \frac{b^{\prime}%
r-3b}{r^{4}}\,\left(  \frac{1}{\omega}+1\right)  -\frac{b}{r^{3}}\frac
{\omega^{\prime}}{\omega^{2}}\right]  \,,\\
U_{2}^{\prime}(r)  &  =-\frac{12}{r^{3}}\left(  1-\frac{b}{r}\right)
+\frac{6(b-b^{\prime}r)}{r^{4}} +\frac{3}{2}\left[  \frac{b^{\prime}%
r-3b}{r^{4}}\,\left(  \frac{1}{3\omega}-1\right)  -\frac{b}{r^{3}}\frac
{\omega^{\prime}}{3\omega^{2}}\right]  \,,
\end{align}
and once evaluated at the throat take the following form
\begin{align}
U_{1}^{\prime}(r_{0})  &  =\frac{3\mathcal{A}}{r_{0}^{3}}\,, \qquad
\mathrm{with} \qquad\mathcal{A}=2\left(  \frac{1+\omega_{0}}{\omega_{0}%
}\right)  -\frac{1}{2\omega_{0}^{2}} \left[  \left(  1+3\omega_{0}\right)
\left(  1+\omega_{0}\right)  +\omega^{\prime}_{0}\,r_{0}\right]
\,,\label{p01}\\
U_{2}^{\prime}(r_{0})  &  =\frac{3\mathcal{B}}{r_{0}^{3}}\,, \qquad
\mathrm{with} \qquad\mathcal{B}=2\left(  \frac{1+\omega_{0}}{\omega_{0}%
}\right)  -\frac{1}{2\omega_{0}^{2}} \left[  \left(  1+3\omega_{0}\right)
\left(  1-3\omega_{0}\right)  +\frac{\omega^{\prime}_{0}\,r_{0}}{3}\right]
\,, \label{p02}%
\end{align}
where the relationship $b^{\prime}=-b/(\omega r)$ has been used to eliminate
the $b^{\prime}$ dependence. We have defined $\omega(r_{0})=\omega_{0}$, and
the term $\omega^{\prime}_{0}\,r_{0}$ is a number as $\omega^{\prime}_{0}
\propto1/r_{0}$. The potentials, Eqs. (\ref{p1})-(\ref{p2}) evaluated at the
throat reduce to
\begin{align}
U_{1}(r_{0})  &  =\frac{3X}{2r_{0}^{2}}\qquad\mathrm{with}\qquad X=\left(
\frac{1+\omega_{0}}{\omega_{0}}\right)  \,,\label{pot1}\\
U_{2}(r_{0})  &  =-\frac{3Y}{2r_{0}^{2}}\qquad\mathrm{with}\qquad Y=\left(
\frac{3\omega_{0}-1}{3\omega_{0}}\right)  \,. \label{pot2}%
\end{align}

To determine the extremum of Eq. (\ref{selfsust4}) with respect to $r$, we
shall consider, for convenience, the following definitions
\begin{equation}
\mathcal{F}(r)=\frac{\omega(r)r^{3}}{b(r)}, \qquad\mathcal{F}^{\prime
}(r)=\frac{r^{2}}{b(r)}[1+3\omega(r)+\omega^{\prime}(r)r]\,,
\end{equation}
where in the latter expression, $b^{\prime}=-b/(\omega r)$ is
used. Thus, the extremum of Eq. (\ref{selfsust4}) with respect to
$r$, takes the form
\begin{equation}
\mathcal{F}^{\prime}(r)\,\left[  U_{1}^{2}\ln\left(  \frac{|U_{1}| \sqrt{e}%
}{4\mu_{0}^{2}} \right)  +U_{2}^{2}\ln\left(  \frac{|U_{2}| \sqrt{e}}{4\mu
_{0}^{2}}\right)  \right]  + \mathcal{F}(r)\,\left[  2U_{1}U_{1}^{\prime}%
\ln\left(  \frac{|U_{1}| e}{4\mu_{0}^{2}} \right)  +2U_{2}U_{2}^{\prime}%
\ln\left(  \frac{|U_{2}| e}{4\mu_{0}^{2}}\right)  \right]  =0 , \label{Ap2}%
\end{equation}
where the following relationship
\begin{equation}
\left[  U_{i}^{2}\ln\left(  \frac{|U_{i}| \sqrt{e}}{4\mu_{0}^{2}} \right)
\right]  ^{\prime}=2U_{i}U_{i}^{\prime}\ln\left(  \frac{|U_{i}| e}{4\mu
_{0}^{2}} \right)  ,
\end{equation}
has been used.

\begin{figure}[h]
\centering
\includegraphics[width=2.8in]{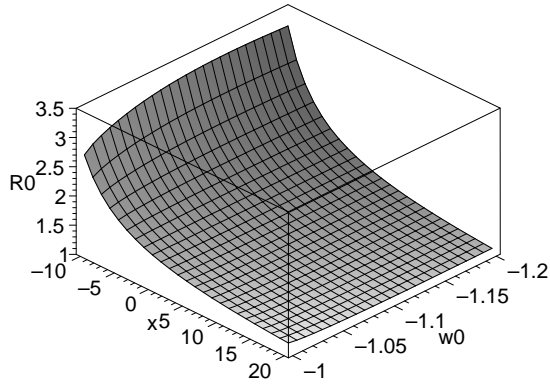} \caption{We have defined the
dimensionless quantity $R_{0}=\bar{r}_{0}/\left(  \sqrt{\frac{3\sqrt{e}}%
{8\mu_{0}^{2}}}\right)  =\sqrt{\Omega}$, which is depicted by the surface.
Note that the throat size decreases for increasing values of the parameter
$x=\omega_{0}^{\prime}r_{0}$.}%
\label{fig:throat}%
\end{figure}

Now, evaluated at the throat, $r_{0}$, in order to have only one solution,
using Eqs. (\ref{p01})-(\ref{p02}), and the expressions for the potentials,
Eqs. (\ref{pot1})-(\ref{pot2}), we verify that Eq. (\ref{Ap2}) provides
\[
\frac{1}{4}(1+3\omega_{0}+\omega_{0}^{\prime}\,r_{0})\left[
X^{2}\,\ln\left( \frac{3\sqrt{e}\,X}{8\mu_{0}^{2}r_{0}^{2}}\right)
+Y^{2}\,\ln\left(
\frac{3\sqrt{e}\,Y}{8\mu_{0}^{2}r_{0}^{2}}\right)  \right]
+\omega_{0}\left[ X\mathcal{A}\,\ln\left(
\frac{3e\,X}{8\mu_{0}^{2}r_{0}^{2}}\right)  -Y{\cal B}\,\ln\left(
\frac{3e\,Y}{8\mu_{0}^{2}r_{0}^{2}}\right) \right] =0\,.
\]
Going through the whole tedious, but fairly straightforward, calculation of
the throat, we arrive at
\begin{equation}
\bar{r}_{0}^{2}=\frac{3\sqrt{e}}{8\mu_{0}^{2}}\,\Omega\,,\qquad\mathrm{with}%
\qquad\Omega=X^{\frac{A+C}{E}}\,Y^{\frac{B+D}{E}}\;\exp\left(  \frac{A+B}%
{2E}\right)  \,. \label{extreme-r0}%
\end{equation}
where the constants are defined as
\begin{align}
A  &  =\omega_{0}\,X\mathcal{A}\,,\qquad B=-\omega_{0}\,Y\mathcal{B}\,,\\
C  &  =\frac{1}{4}\,(1+3\omega_{0}+\omega_{0}^{\prime}\,r_{0})X^{2}\,,\\
D  &  =\frac{1}{4}\,(1+3\omega_{0}+\omega_{0}^{\prime}\,r_{0})Y^{2}\,,
\end{align}
and $E=A+B+C+D$. We may now consider $\omega_{0}<-1$ (phantom
regime) and $\omega_{0}^{\prime}\,r_{0}$ as free parameters, and
consequently plot $r_{0}$, which is depicted in Fig.
\ref{fig:throat}. We have defined the
dimensionless quantity $R_{0}=\bar{r}_{0}/\left(  \sqrt{\frac{3\sqrt{e}}%
{8\mu_{0}^{2}}}\right)  =\sqrt{\Omega}$, which is depicted by the surface in
Fig. \ref{fig:throat}. Note that the throat size decreases for increasing
values of the parameter $x=\omega_{0}^{\prime}r_{0}$. 

\begin{figure}[h] \centering
\includegraphics[width=2.8in]{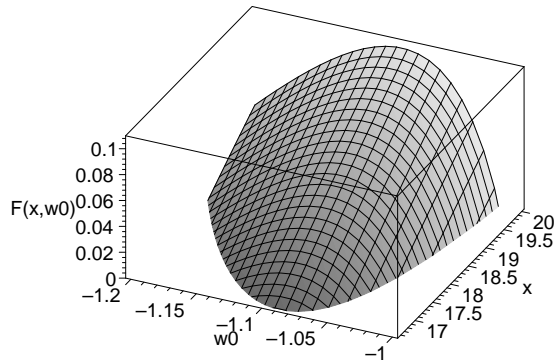} \caption{The log in the definition of
$G_{0}(\mu_{0})$ must be positive, which places restrictions on the range of
$x=\omega^{\prime}_{0} r_{0}$. The surface in the figure is given by
$F(x,\omega_{0})=\ln[(X/\Omega)^{X^{2}}(Y/\Omega)^{Y^{2}}]$. Only solutions
with a steep positive slope of the parameter $\omega_{0}^{\prime}r_{0}$, at
the throat, are permitted. As $\omega_{0}$ tends to $-1$, we verify that
$\omega_{0}^{\prime}r_{0}$ increases significantly. Note that solutions with a
constant parameter $\omega$, i.e., $\omega'_{0}=0$, are prohibited.}%
\label{fig:G0}%
\end{figure}

Now, restrictions on the parameter range of $\omega_{0}$ and $\omega
_{0}^{\prime}r_{0}$ may also be imposed. The quantity $G_{0}(\mu_{0})$ must be
positive, so that Eq. (\ref{selfsust4}) evaluated at the throat, reduces to
\begin{equation}
\frac{1}{G_{0}(\mu_{0})}=\frac{3|\omega_{0}|\,\mu_{0}^{2}}{8\pi^{2}\sqrt
{e}\,\Omega}\, \ln\left[  \left(  \frac{X}{\Omega}\right)  ^{X^{2}}\left(
\frac{Y}{\Omega}\right)  ^{Y^{2}}\right]  . \label{selfsust5}%
\end{equation}
The log must be positive, which places restrictions on the range of
$\omega^{\prime}_{0} r_{0}$. We shall depict $F=\ln[ (X/\Omega)^{X^{2}%
}(Y/\Omega)^{Y^{2}}] $ in Fig. \ref{fig:G0}. Note that only
solutions with a steep positive slope proportional to
$\omega_{0}^{\prime}$, at the throat, are permitted. As
$\omega_{0}\rightarrow-1$, we verify that
$\omega_{0}^{\prime}r_{0}$ increases significantly. Solutions with
a constant parameter $\omega$, i.e., $\omega'_{0}=0$, are
prohibited, which is in agreement with the calculations in
Appendix \ref{app}.

An interesting generic constraint on $\mu/\mu_{0}$ may also be deduced.
Substituting Eqs. (\ref{extreme-r0}) and (\ref{selfsust5}) into Eq.
(\ref{selfsust4a}), evaluated at the throat, provides the following
relationship
\begin{equation}
\frac{1}{G_{0}(\mu)}=\frac{3|\omega_{0}|\,\mu_{0}^{2}}{8\pi^{2}\sqrt
{e}\,\Omega}\,\ln\left[  \left(  \frac{X}{\Omega}\right)  ^{X^{2}}\left(
\frac{Y}{\Omega}\right)  ^{Y^{2}}\Big/\left(  \frac{\mu}{\mu_{0}}\right)
^{2(X^{2}+Y^{2})}\right]  \,. \label{selfsust4b}%
\end{equation}
Now $G_{0}(\mu)$ must be positive, so that the above relationship places a
constraint on $\mu/\mu_{0}$, namely
\begin{equation}
\mu<\mu_{0}\left[  \left(  \frac{X}{\Omega}\right)  ^{X^{2}}\left(  \frac
{Y}{\Omega}\right)  ^{Y^{2}}\right]  ^{\frac{1}{2(X^{2}+Y^{2})}}\,.
\end{equation}

The above results have been based on a variational calculation,
concerning the one-loop graviton contribution in a wormhole
background. On general grounds, in order for the one-loop
contribution to stabilize the wormhole it can only do so at the
Planck scale, so that the wormhole throat is expected to be of the
Planck order. We have found an upper bound on the mass parameter
$\mu$, and we emphasize that we are not interested in computing an
exact result of the latter, but in the possibility that phantom
wormholes be, in principle, sustained by their own quantum
fluctuations. Indeed, we have demonstrated this fact in the
context of an $r$-dependent equation of state parameter within the
one-loop graviton contribution approach.

\section{Conclusion}

\label{sec:conlusion}

Inspired by the evolving dark energy parameter crossing the
phantom divide, we have considered in this work a varying equation
of state parameter dependent on the radial coordinate, i.e.,
$\omega(r)=p(r)/\rho(r)$, and imposed that phantom energy is
concentrated in the neighborhood of the throat. The pressure in
the equation of state is a negative radial pressure, and the
tangential pressure $p_t$ is determined through the Einstein field
equations. We have also considered the possibility that these
phantom wormholes be sustained by their own quantum fluctuations.
The energy density of the graviton one loop contribution to a
classical energy in a phantom wormhole background, and the finite
one loop energy was considered as a self-consistent source for
these phantom wormholes. The size of the wormhole throat as a
function of the parameters $\omega_{0}$ and $\omega'_{0}r_0$ was
further explored, and in order for the one-loop contribution to
stabilize the wormhole it can only do so at the Planck scale, so
that the wormhole throat is expected to be of the Planck order.
In a rather speculative scenario, one may also argue that the
accretion of phantom energy by the wormhole gradually increases
the wormhole throat to macroscopic size, much in the spirit of
Refs. \cite{gonzalez2,diaz-phantom3}.

At first sight one can think that the \textit{phantom} region is
straightforward to analyze. However, within the approach of Ref.
\cite{Garattini}, this is not trivial. Indeed, the main difference
is in the integration. In this paper, we have worked with the
energy density, while in Ref. \cite{Garattini}, the integration
over the whole space was considered. In Appendix \ref{app}, we
show that the classical term integrated over all space forbids the
appearance of \textit{phantom} energy for constant $\omega$
\cite{Garattini1}.
However, it is important to emphasize that the semi-classical
approach used in this work only rules out self sustained
wormholes, with constant $\omega$, concerning the one-loop
graviton contribution, and not the whole plethora of phantom
wormholes, as the ones analyzed in Refs.
\cite{phantomWH,phantomWH2}.
A few words should be spent concerning the difference between the
\textit{phantom} and the \textit{dark} energy case. Actually the
\textit{dark} energy (not energy density) could be considered,
because the classical term strongly forbids \textit{phantom}
energy. Nevertheless, the \textit{dark} energy region is
completely incompatible with asymptotic
flatness~\cite{Garattini1}. The analysis for $\omega(r)$ is not a
trivial task and it depends on a case to case. However, we have
shown that interesting solutions do, in fact, exist, namely those
that permit solutions with a steep positive slope proportional to
the radial derivative of the equation of state parameter,
evaluated at the throat.
It is rather important to emphasize a shortcoming in the analysis
carried in this paper, mainly due to the difficulties of the
semi-classical theory of gravity and, in particular to the
technical problems encountered. We have considered a
semi-classical variational approach using the energy density,
which imposes a local analysis to the problem. Thus, we have
restricted our attention to the behavior of the metric function
$b(r)$, through a variational approach, at the wormhole throat,
$r_0$. Despite the fact that the behavior is unknown far from the
throat, due to the high curvature effects at or near $r_0$, the
analysis carried out in this work should extend to the immediate
neighborhood of the wormhole throat.
Therefore, we may conclude that, in principle, phantom wormhole
throats with a varying equation of state parameter dependent on
the radial coordinate, may be self sustained by their quantum
fluctuations within the semi-classical approach outlined in this
paper.

%-----------------------------------
\section*{Acknowledgements}
%-----------------------------------

FSNL was funded by Funda\c{c}\~{a}o para a Ci\^{e}ncia e
Tecnologia (FCT)--Portugal through the research grant
SFRH/BPD/26269/2006.

\appendix
%-----------------------------------------------

\section{Computation of the classical term}

%-----------------------------------------------
\label{app}
%-----------------------------------------------

Here, we compute the hamiltonian of Eq. (\ref{classical}) when $\omega(r)$ is
a constant. It is easy to show that%
\begin{equation}
H_{\Sigma}^{(0)}=\frac{1}{G\omega}\int_{r_{0}}^{+\infty}dr\left(  \frac{r_{0}%
}{r}\right)  ^{\frac{x}{2}}\frac{1}{\sqrt{\left(  \frac{r}{r_{0}}\right)
^{x}-1}}\qquad x:=1+\frac{1}{\omega}.
\end{equation}
In the previous integral there is an extra factor \textquotedblleft%
2\textquotedblright\ coming from the counting of the universes. We change
variable to obtain%
\begin{equation}
\frac{2r_{0}}{G\omega x}\int_{0}^{+\infty}\frac{dt}{\cosh^{2-2/x}\left(
t\right)  }=\frac{r_{0}}{G\left(  1+\omega\right)  }B\left(  \frac{1}{2}%
,\frac{1}{1+\omega}\right)  =\frac{r_{0}}{G}A\left(  \omega\right)
\label{intE}%
\end{equation}
with%
\begin{equation}
A\left(  \omega\right)  =\frac{1}{\left(  1+\omega\right)  }B\left(  \frac
{1}{2},\frac{1}{1+\omega}\right)  =\frac{\sqrt{\pi}}{\left(  1+\omega\right)
}\frac{\Gamma\left(  \frac{1}{1+\omega}\right)  }{\Gamma\left(  \frac
{3+\omega}{2+2\omega}\right)  }.
\end{equation}
In Eq.$\left(  \ref{intE}\right)  $, we have used the following formula%
\begin{equation}
\int_{0}^{+\infty}dt\frac{\sinh^{\mu}t}{\cosh^{\nu}t}=\frac{1}{2}B\left(
\frac{\mu+1}{2},\frac{\nu-\mu}{2}\right)  \qquad\left\{
\begin{array}
[c]{c}%
\operatorname{Re}\mu>-1\\
\operatorname{Re}\left(  \mu-\nu\right)  <0
\end{array}
\right.  .
\end{equation}
The existence of the integral is submitted to the following condition
$(\mu=0)$ $\omega>-1$ and this shows that the phantom region is forbidden.

\end{document}